\documentclass[12pt]{article}
\usepackage{mathrsfs}
\usepackage[T1]{fontenc}
\usepackage{mathpazo}
\usepackage{setspace}
\usepackage{amsfonts}
\usepackage{amssymb}
\usepackage{amsmath}
\usepackage{esint}                   % For closed integrals
\usepackage{epsfig}
\usepackage{latexsym}
\usepackage{color}
\usepackage{graphicx}
\usepackage{hyperref}
\usepackage{nicefrac}
\usepackage[latin1]{inputenc}
\usepackage{pstricks}
\usepackage{slashed}
\usepackage{multirow}
\usepackage{ulem}
\usepackage{comment}
\usepackage[nosort]{cite}
\usepackage{datetime}
\usepackage{tikz-cd}
\usepackage{float}

\usepackage[greek,english]{babel}
\languageattribute{greek}{polutoniko}

\usepackage[vcentermath]{youngtab}

\def\hybrid{\topmargin -30pt    \oddsidemargin 0pt %%%%%%%%%%%%%% Archive-30pt
        \headheight 0pt \headsep 0pt
        \textwidth 6.25in       % A4 paper
        \textheight 9.5in       % A4 paper
        \marginparwidth .875in
        \parskip 5pt plus 1pt   \jot = 1.5ex}

\hybrid

\def\baselinestretch{1.2}

\catcode`\@=11

\def\marginnote#1{}
\newcount\hour
\newcount\minute
\newtoks\amorpm
\hour=\time\divide\hour by60
\minute=\time{\multiply\hour by60 \global\advance\minute by-\hour}
\edef\standardtime{{\ifnum\hour<12 \global\amorpm={am}%
        \else\global\amorpm={pm}\advance\hour by-12 \fi
        \ifnum\hour=0 \hour=12 \fi
        \number\hour:\ifnum\minute<10 0\fi\number\minute\the\amorpm}}
\edef\militarytime{\number\hour:\ifnum\minute<10 0\fi\number\minute}
\def\draftlabel#1{{\@bsphack\if@filesw {\let\thepage\relax
   \xdef\@gtempa{\write\@auxout{\string
      \newlabel{#1}{{\@currentlabel}{\thepage}}}}}\@gtempa
   \if@nobreak \ifvmode\nobreak\fi\fi\fi\@esphack}
        \gdef\@eqnlabel{#1}}
\def\@eqnlabel{}
\def\@vacuum{}
\def\draftmarginnote#1{\marginpar{\raggedright\scriptsize\tt#1}}

\def\draft{\oddsidemargin -.5truein
        \def\@oddfoot{\sl preliminary draft \hfil
        \rm\thepage\hfil\sl\today\quad\militarytime}
        \let\@evenfoot\@oddfoot \overfullrule 3pt
        \let\label=\draftlabel
        \let\marginnote=\draftmarginnote
   \def\@eqnnum{(\theequation)\rlap{\kern\marginparsep\tt\@eqnlabel}%
\global\let\@eqnlabel\@vacuum}  }

\def\draft2{
        \def\@oddfoot{\sl preliminary draft \hfil
        \rm\thepage\hfil\sl\today\quad\militarytime}
        \let\@evenfoot\@oddfoot \overfullrule 3pt
        \let\marginnote=\draftmarginnote
   \def\@eqnnum{(\theequation)\rlap{\kern\marginparsep\tt\@eqnlabel}%
\global\let\@eqnlabel\@vacuum}  }

\def\preprint{\twocolumn\sloppy\flushbottom\parindent 2em
        \leftmargini 2em\leftmarginv .5em\leftmarginvi .5em
        \oddsidemargin -.5in    \evensidemargin -.5in
        \columnsep .4in \footheight 0pt
        \textwidth 10.in        \topmargin  -.4in
        \headheight 12pt \topskip .4in
        \textheight 6.9in \footskip 0pt
        \def\@oddhead{\thepage\hfil\addtocounter{page}{1}\thepage}
        \let\@evenhead\@oddhead \def\@oddfoot{} \def\@evenfoot{} }

\def\numberbysection{\@addtoreset{equation}{section}
        \def\theequation{\thesection.\arabic{equation}}}

\def\underline#1{\relax\ifmmode\@@underline#1\else
        $\@@underline{\hbox{#1}}$\relax\fi}

\def\titlepage{\@restonecolfalse\if@twocolumn\@restonecoltrue\onecolumn
     \else \newpage \fi \thispagestyle{empty}\c@page\z@
        \def\thefootnote{\fnsymbol{footnote}} }

\def\endtitlepage{\if@restonecol\twocolumn \else \newpage \fi
        \def\thefootnote{\arabic{footnote}}
        \setcounter{footnote}{0}}  %\c@footnote\z@ }

\catcode`@=12
\relax

\def\figcap{\section*{Figure Captions\markboth
        {FIGURECAPTIONS}{FIGURECAPTIONS}}\list
        {Figure \arabic{enumi}:\hfill}{\settowidth\labelwidth{Figure
999:}
        \leftmargin\labelwidth
        \advance\leftmargin\labelsep\usecounter{enumi}}}
 \relax
\def\tablecap{\section*{Table Captions\markboth
        {TABLECAPTIONS}{TABLECAPTIONS}}\list
        {Table \arabic{enumi}:\hfill}{\settowidth\labelwidth{Table
999:}
        \leftmargin\labelwidth
        \advance\leftmargin\labelsep\usecounter{enumi}}}
 \relax
\def\reflist{\section*{References\markboth
        {REFLIST}{REFLIST}}\list
        {[\arabic{enumi}]\hfill}{\settowidth\labelwidth{[999]}
        \leftmargin\labelwidth
        \advance\leftmargin\labelsep\usecounter{enumi}}}
 \relax
\makeatletter
\newcounter{pubctr}
\def\publist{\@ifnextchar[{\@publist}{\@@publist}}
\def\@publist[#1]{\list
        {[\arabic{pubctr}]\hfill}{\settowidth\labelwidth{[999]}
        \leftmargin\labelwidth
        \advance\leftmargin\labelsep
        \@nmbrlisttrue\def\@listctr{pubctr}
        \setcounter{pubctr}{#1}\addtocounter{pubctr}{-1}}}
\def\@@publist{\list
        {[\arabic{pubctr}]\hfill}{\settowidth\labelwidth{[999]}
        \leftmargin\labelwidth
        \advance\leftmargin\labelsep
        \@nmbrlisttrue\def\@listctr{pubctr}}}
 \relax
\makeatother

\def\be{\begin{equation}}
\def\ee{\end{equation}}
\def\ba{\begin{eqnarray}}
\def\ea{\end{eqnarray}}

\def\del{\partial}

\def\a{\alpha}

\def\b{\beta}

\def\g{\gamma}

\def\d{\delta}
\def\D{\Delta}

\def\bk{{\bf k}}
\def\bz{{\bf z}}

\def\no{\noindent}

\def\qq{\qquad}

\def\IR{\relax{\rm I\kern-.18em R}}

\def\inv{^{\raise.0ex\hbox{${\scriptscriptstyle -}$}\kern-.05em 1}}

\def \ha {{\frac{1}{2}}}

\def \ov {\over}

\def\diag{{\rm diag}}

\def\half{{\textstyle {1 \over 2}}}

\newcommand{\bb}{\hskip -0.1cm}

\newcommand{\hm}{\hskip -0.05cm - \hskip -0.05cm}

\def\tr{\textrm{Tr}}

\begin{document}
%\draft2

%\renewcommand{\theequation}{\arabic{equation}}
%\renewcommand{\theequation}{\thesection.\arabic{equation}}

\renewcommand{\theequation}{\thesection.\arabic{equation}}
\csname @addtoreset\endcsname{equation}{section}

\begin{titlepage}
\begin{center}
\hfill CERN-TH-2023-085

\renewcommand*{\thefootnote}{\arabic{footnote}}

\phantom{xx}
\vskip 0.5in

{\large {\bf Composing arbitrarily many $SU(N)$ fundamentals}}

\vskip 0.4in

{\bf Alexios P. Polychronakos$^{1,2}$}\hskip .15cm and \hskip .15cm
{\bf Konstantinos Sfetsos}$^{3,4}$

\vskip 0.14in

${}^1\!$ Physics Department, the City College of New York\\
160 Convent Avenue, New York, NY 10031, USA\\
{\footnotesize{\tt apolychronakos@ccny.cuny.edu}}\\
\vskip 0.3cm
${}^2\!$ The Graduate School and University Center, City University of New York\\
365 Fifth Avenue, New York, NY 10016, USA\\
{\footnotesize{\tt apolychronakos@gc.cuny.edu}}

\vskip .14in

${}^3\!$
Department of Nuclear and Particle Physics, \\
Faculty of Physics, National and Kapodistrian University of Athens, \\
Athens 15784, Greece\\
{\footnotesize{ ksfetsos@phys.uoa.gr}}\\

\vskip .14in
${}^4\!$ Theoretical Physics Department,
 CERN, \\1211 Geneva 23, Switzerland

\vskip .3in
\today

\vskip .2in

\end{center}

\vskip .2in

\centerline{\bf Abstract}

\no
We compute the multiplicity of the irreducible representations in the decomposition of the tensor
product of an arbitrary number $n$ of fundamental representations of $SU(N)$, and we identify a duality
in the representation content of this decomposition. Our method utilizes the
mapping of the representations of $SU(N)$ to the states of free fermions on the circle, and can be
viewed as a random walk on a multidimensional lattice. We also derive the large-$n$ limit and the response
of the system to an external non-abelian magnetic field. These results can be used to study the phase
properties of non-abelian ferromagnets and to take various scaling limits. 
\vskip .4in

\vfill

\end{titlepage}
\vfill
\eject

%\def\baselinestretch{1.2}
%\baselineskip 10 pt
%\noindent

%\tableofcontents

\def\baselinestretch{1.2}
\baselineskip 20 pt

\newcommand{\eqn}[1]{(\ref{#1})}

\tableofcontents

%%%%%%%%%%%%%

\section{Introduction}
\label{intro}

Symmetries in physics and in mathematics are often encoded in unitary groups. In quantum mechanics,
in particular, many physical systems realize representations of unitary groups.
The canonical example is spin, an irreducible representation of the group $SU(2)$ (a double
cover of the group of spatial rotations $SO(3)$), with isospin (also $SU(2)$) and elementary particle
"flavor" symmetry ($SU(3)$) constituting more exotic examples.
Systems consisting of several components, each carrying an irreducible representation of a symmetry group,
transform under the same symmetry, in the direct sum of the representations of the components.
The decomposition of the states of the full system into
irreducible representations of the symmetry group becomes, then, a problem of physical relevance.

The group theoretical techniques of decomposing a direct sum of two irreducible representations (irreps) are
well established \cite{Ful}. In particular, the irrep content of the decomposition of the tensor product
of irreps of $SU(N)$ has been studied in the mathematics literature \cite{kir,kul}. The result for the tensor
product of fundamental irreps, in particular, can be derived from Schur-Weyl duality
\cite{FuHa}.\footnote{We thank Jules Lamers for pointing out this connection.}
The existing derivations, however, use a rather
heavily mathematical language and the results are in a somewhat implicit form. A derivation in a physically
motivated language and in a way accessible to physicists was desirable.
In \cite{Polychronakos:2016vpb} we undertook this task for the simplest case of  $SU(2)$.
Specifically, we computed the multiplicity of spin $j$ irreps arising from composing $n$ spins $s$
(related results were also derived in \cite{Men,Curtright:2016eni} and used in \cite{Raf}, and were
further elaborated in \cite{Gyamfi:2018crc}).
We also derived explicit expressions for
the multiplicities arising from spins corresponding to indistinguishable particles by distilling
the totally symmetric (for bosons) or antisymmetric (for fermions) parts of the decomposition.
We could recast the composition problem as random walks on a one-dimensional lattice,
or equivalently generalized Dyck or Lukasiewicz paths \cite{Leh} with appropriate boundary conditions.
Even for the $SU(2)$ case the results are interesting, exhibiting nontrivial scaling properties
in the large-$n$ and large-$s$ limits, especially for bosonic or fermionic spins.
In \cite{Polychronakos:2016vpb} we provided the asymptotic expressions in these limits, and explored
phase transitions occuring in a system of interacting spins, possibly coupled to an external
magnetic field.

An interesting question is whether the general $SU(N)$ case manifests similar phenomena and if it
has qualitatively different properties. Investigating it requires having explicit expressions for
the multiplicities in the decomposition of the direct sum of an arbitrary number of $SU(N)$
irreps. Since representations of $SU(N)$ are in general labeled by $N-1$ nonnegative integers (the Young
tableau row lengths), the corresponding general formulas for the multiplicities are in general more 
complicated and implicit. In the present work we simplify somewhat the problem by restricting to the
direct sum of
$n$ fundamental representations of $SU(N)$. As we shall see, this problem still has a rich structure.

Our motivation for carrying out this analysis is manifold. Firstly, we wished to present a complete and
explicit solution to the problem. Moreover, we wanted to present an intuitive and pedagogical approach,
building on
physics connections and intuition. In this vein, we used the fact that this problem can be related to a
system of $N$ free quantum mechanical fermions on a circle. Finally, systems of many $SU(N)$
degrees of freedom arise in condensed matter and ultracold systems\cite{Gor,Kap}, 
spin chains \cite{Aff,Pola}, particle physics \cite{Hoo,BIPZ}, matrix models \cite{GrMi},
etc., and we wanted to pave the road for possible physical applications.

In this paper we present a full solution to the problem of decomposing the direct sum of an arbitrary
number of $SU(N)$ irreps and apply it to the case of fundamental irreps, deriving explicit formulae for
the representation content of the decomposition. The example of $SU(3)$ is analyzed in some detail.
We point out a random walk connection, this time on a $N-1$-dimensional lattice, and identify a
duality relation in the distribution of irreps. We also consider the limit of a large number $n$ of
fundamentals, as a prelude to more general and intricate limiting cases.
Finally, we calculate the response of the system to coupling it with a nonabelian magnetic
field, which is relevant in the study of the thermodynamics of interacting $SU(N)$
systems. The structure and phase transitions of the thermodynamic limit of such systems are treated
in \cite{Phases}, and nontrivial scaling limits involving both $N$ and $n$ taken large will be investigated
in forthcoming publications. 

We present our results in the upcoming sections, hinting at additional physical applications and possible
extensions of our analysis in the conclusions.

%%%%%%%%%%%
%
%\section{Product of many irreps of $SU(N)$}
%\label{generalities}

\section{Fermion representation of $SU(N)$ group theory}
\label{generalities}

We review here the description of irreducible representations (irreps) of $SU(N)$ as
$N$-fermion energy eigenstates on the circle \cite{MiPo} and the corresponding composition rules in the
fermion picture. This representation affords a conceptual standpoint
that is well-suited to our considerations and will be useful throughout the rest of the paper. It also
leads naturally to a duality relation between groups and representations.

\subsection{Quantum mechanics on the $U(N)$ manifold}

The correspondence of irreps of $SU(N)$ and $N$-fermion energy eigenstates can be most naturally
established by considering the action of a particle moving freely on the group manifold $U(N)$ \cite{Polb}.
The Lagrangian is the kinetic energy of the particle
\be
{\cal L} = -\half \tr \left( U^{-1} {\dot U} \right)^2\ ,
\label{Lsun}
\ee
where overdot signifies time derivative. This is essentially a unitary matrix model.

One approach to find the energy eigenstates of the corresponding
Hamiltonian is to note that, upon defining momenta and quantizing, the Hermitian matrix operators
$L = i {\dot U} U^{-1}$ and $R = -i U^{-1} {\dot U} =- U^{-1} L U$ 
satisfy the $U(N)$ algebra and mutually commute 
\be
[ L_{ij} , L_{kl} ] = {\rm i} L_{il} \delta_{kj} - {\rm i} L_{kj} \delta_{il}\, ,\quad 
[ R_{ij} , R_{kl} ] = {\rm i} R_{il} \delta_{kj} - {\rm i} R_{kj} \delta_{il}\, ,\quad [ L_{ij} , R_{kl} ] = 0\ .
\ee
Left and right multiplications of $U$ by unitary matrices are generated by $L$ and $R$, respectively,
which conforms with their commutation relations. $L$ and $R$ have equal and opposite $U(1)$ parts
$\sum_k L_{kk} = -\sum_k R_{kk}$ and share a common quadratic Casimir. 
They are both constants of motion, and the Hamiltonian is their second Casimir. Energy eigenstates are matrix elements of $U$ in arbitrary $U(N)$ irreps $r$:
$r_{ab} (U)$, $a,b = 1,\dots,d_r$, transforming in the $r \otimes {\bar r}$ representation under
$L \otimes R$, with energy eigenvalue $C_2 (r)$ and degeneracy $d_r^2$.

Now impose the additional first-class constraint $L+R =0$, which commutes with the Hamiltonian. It
amounts to choosing states that are invariant under the combined left and right unitary rotations
$U \to V U V^{-1}$ generated by $L+R$, that is, under unitary conjugations. Energy eigenstates become
the conjugation-invariant linear combinations $\sum_a r_{aa} (U) = \tr_r (U) = \chi_r (U)$, the latter
being the  character of the irrep.
We recover the same spectrum,
with states corresponding to irreps $r$ of $SU(N)$, but now with degeneracy one due to the trace.

The condition $L+R=0$, or, classically, the vanishing of the matrix commutator $[U , \dot U] = 0$,
implies (using the equation of motion for $U$) that $U(t)$ and $U(0)$ commute as
matrices. Hence, $U(t)$ can be diagonalized by a time-independent unitary conjugation. Implementing
this at the level of the Lagrangian by writing
%\be
%U(t) = V\, \diag \{e^{i x_j (t)} \}\, V^{-1} \, ,\quad  x_j \sim x_j + 2\pi\ ,
%\ee
\be
U(t) = V\, \diag \{ z_j \}\, V^{-1} \, ,\quad z_j := e^{i x_j} \, ,~~~ x_j \sim x_j + 2\pi\ ,
\ee
we see that (\ref{Lsun}) becomes the Lagrangian of $N$ free particles on the unit circle with coordinates
$x_j$. Since exchanging the eigenvalues is a special case of unitary conjugation, states
$\phi (x_1,\dots,x_N)$ are invariant under exchange of the $x_j$ and are, in principle, bosonic.
However, the change of variables from $U$ to $x_j$ introduces the standard measure, equal to the
absolute square of the Vandermonde determinant $|\Delta|^2$. In terms of the exponential variables $\bz$
(boldfaced $\bz$ stands for the full set of variable $z_1, \dots , z_N$, and similarly for
other sets of $N$ quantities), $\Delta(\bz)$ is given by
\be
\label{vanderm}
\Delta (\bz) = (z_1 \cdots z_N )^{-{N-1 \over 2}} \left|
\begin{array}{ccccccccc}
z_1^{N-1} & z_1^{N-2} & \cdots & z_1  & 1 \\
z_2^{N-1} & z_2^{N-2} & \cdots & z_2  & 1 \\
\vdots & \vdots & \ddots & \vdots & \vdots  \\
z_N^{N-1} & z_N^{N-2} & \cdots & z_N  & 1 \\
\end{array}
\right|
=\prod_{i>j=1}^N 2 {\rm i} \sin{x_i - x_j \over 2}\ .
\ee
The prefactor $ (z_1 \cdots z_N )^{-{N-1 \over 2}}$ is a pure phase introduced to make
$\Delta (\bz)$ invariant under uniform shifts of the $x_i$'s.

\no
Upon incorporating one factor $\Delta (\bz)$ into the wavefunction
\be
\psi (\bz) = \Delta (\bz)\, \phi (\bz)\ ,
\label{psen}\ee
the Hamiltonian becomes the standard free $N$-particle Hamiltonian on the $x_j$ (compare with the
transformation $u(r) = r \phi (r)$ in the Schr\"odinger equation in spherical coordinates). Because
of the prefactor $\Delta (\bz)$ the states $\psi (\bz)$ are  antisymmetric upon 
interchanging any two of the $x_i$. This establishes
the correspondence between the singlet sector of the model (\ref{Lsun}) and free fermions on the
circle, and thus of irreps of $SU(N)$ and free fermion energy eigenstates. In particular, the fermion
energy eigenstate corresponding to an irrep $r$ and energy $E = C_2 (r)$ is
\be
\psi_r (\bz) = \Delta (\bz) \,\chi_r (\bz)\ ,
\label{psar}\ee
with $\chi_r (\bz) = \tr\, r(U)$ the character of irrep $r$ in terms of the eigenvalues $z_j$ of $U$.

\subsection{Correspondence with Young tableaux}

It will be useful to establish the correspondence of fermion states and Young tableaux of $SU(N)$.
The single-particle
spectrum on the circle consists of discrete momentum eigenstates with eigenvalue 
$k = 0,\pm 1,\pm 2,\dots$ and energy $E_k = k^2 /2$. An $N$-fermion energy eigenstate
corresponds to filling $n$ of the single-particle states with fermions. Call $k_1 > k_2 > \dots > k_N$
the momenta of these states in decreasing order (see figure 1):
\vskip 0.3cm
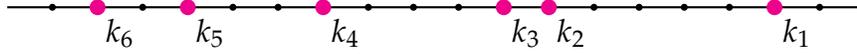
\begin{figure}[h!]
 \centering
 \begin{tikzpicture}[scale=0.6]\hskip 0.3cm
\draw[thick,-] (-1,0) -- (18,0);
\fill (0,0) circle (2.2pt); \fill (1,0)node[below] {$k_6$}[magenta] circle (5pt);\fill (2,0) circle (2.2pt);
\fill (3,0)node[below] {$k_5$}[magenta] circle (5pt);\fill (4,0) circle (2.2pt);\fill (5,0) circle (2.2pt);
\fill (6,0)node[below] {$k_4$}[magenta] circle (5pt);
\fill (7,0) circle (2.2pt);
\fill (8,0) circle (2.2pt);\fill (10,0)node[below] {$k_3$}[magenta] circle (5pt);\fill (12,0) circle (2.2pt);\fill (9,0) circle (2.2pt);
\fill (11,0)node[below] {$k_2$}[magenta] circle (5pt);\fill (16,0)node[below] {$k_1$}[magenta] circle (5pt);\fill (13,0) circle (2.2pt);\fill (14,0) circle (2.2pt);
\fill (15,0) circle (2.2pt);\fill (17,0) circle (2.2pt);
\end{tikzpicture}
\vskip - 0.3 cm
  \caption{\small The state represents an $SU(6)$ irrep with $\ell_1 =10,\, \ell_2 =\ell_3 =6,\, \ell_4 = 3,\, \ell_5=1$.
  The corresponding Young tableau has five lines with $10, 6, 6, 3$ and $1$ box, respectively.}
\label{figdots}
\end{figure}
\no
The (unnormalized) wavefunction corresponding to this state is given by the Slater determinant
\be
\psi_\bk (\bz) = \left|
\begin{array}{ccccccccc}
z_1^{k_1} & z_1^{k_2} & \cdots & z_1^{k_{N-1}}  & z_1^{k_N} \\
z_2^{k_1} & z_2^{k_2} & \cdots & z_2^{k_{N-1}}  & z_2^{k_N} \\
\vdots & \vdots & \ddots & \vdots & \vdots  \\
z_N^{k_1} & z_N^{k_2} & \cdots & z_N^{k_{N-1}} & z_N^{k_N} 
\end{array}
\right|\ .
\label{slater}
\ee
The total momentum of the fermions $k = k_1 + \cdots + k_n$ corresponds to the
states picking up a phase $e^{i c k}$ upon the shift $x_j \to x_j +c$, that is, upon $U \to e^{ic} U$.
It thus represents the $U(1)$ charge of the state. We may shift all momenta by a constant,
changing $k$ and the $U(1)$ charge without affecting the $SU(N)$ part of the
states, which  can then be labeled by the $N\hm1$ shift-invariant integers
$k_1 -k_N > k_2 -k_N > \dots > k_{N-1} - k_N >0$.
Alternatively, we can neutralize the $U(1)$ charge by introducing the prefactor $(z_1 \dots
z_N)^{-\sum_i k_i /N}$ in \eqn{slater}, similarly to the prefactor introduced in $\Delta ({\bf z})$.
With this additional prefactor, \eqn{slater}  maps to \eqn{vanderm} for the
singlet representation for which $k_i = N-i\,,  i=1, \dots ,N$.

\noindent
The final step involves expressing $k_j -k_N$ in terms of new "bosonic" variables $\ell_j$ as
\be
\label{ellk}
\ell_j = k_j - k_N + j-N\ , \qq \ell_1 \geqslant \ell_2\geqslant\dots \geqslant \ell_{N-1} \geqslant 0\ .
\ee
The non-negative, ordered integers $\ell_j$ represent the lengh of rows $j=1,2,\dots , N\hm1$ of the
Young tableau of the irrep corresponding to the fermionic state. The condition 
$\ell_N =0$ is the statement that Young tableau columns of length $N$ give a singlet, contributing at
most to the $U(1)$ charge, and are eliminated. Fig. \ref{figdots} depicts an example of a fermionic
state equivalent to a Young tableau with five rows. As stated, the total momentum and energy of the
fermion state map to the $U(1)$ charge and the quadratic Casimir of the representation. 

\no
The transition from $k_j$ to $\ell_j$ is, in fact, bosonization, in the sense that two or more $\ell_i$'s can be equal.
 For reference, we also record the different parametrization
\be
\label{mdef}
m_j = \ell_j - \ell_{j+1} = k_j - k_{j+1} -1 \geqslant 0\ ,\qquad j=1,2,\dots,N-1\ .
\ee
The non-negative integers $m_j$ are unrestricted and thus provide a "field theoretical" bosonization of the
system. They can be thought of as corresponding to excitations in the first $N-1$ positive modes $j$ of a
second quantized boson field $\phi (x)$.

\section{Composition of representations}

We will now present the method of composing irreps  and derive explicit expressions for the case of composition of fundamentals. 
We will also work out explicitly some examples of low rank groups.

\subsection{General method}

The composition of representations, and their decomposition into irreps, becomes particularly
convenient in the fermion picture.

\no
Consider the direct product $r_1 \times r_2$ of two (possibly reducible) representations $r_1$ and $r_2$. The basic relation
\be
\tr\, (r_1 \times r_2 ) (U) = \tr\, r_1 (U)\, \tr\, r_2 (U) = \chi_{r_1} (U)\, \chi_{r_2} (U)\ ,
\ee
implies, through  (\ref{psen} and \ref{psar}), that their corresponding fermion states are related as
\be
\label{psis}
\psi_{r_1 \times r_2} (\bz) =\Delta(\bz) \chi_{r_1} (\bz)\chi_{r_2} (\bz)=
 {\psi_{r_1} (\bz)\, \psi_{r_2} (\bz) \over \Delta(\bz)}
= \psi_{r_1} (\bz)\, \chi_{r_2} (\bz) = \psi_{r_2} (\bz)\, \chi_{r_1} (\bz)\ .
\ee
The advantage of the fermionic representation is that from the wavefunction $\psi (\bz)$ we can simply
read off the irreducible components by expanding in monomials $z_1^{k_1} \cdots z_N^{k_N}$ and
mapping them to Young tableaux. This will be demonstrated
in detail in the upcoming sections for the composition of a large number of irreps. We record here the characters
for the fundamental $f$, doubly symmetric (a single row with two boxes) $s$, doubly antisymmetric
(two rows with one box each) $a$, and adjoint $ad$ irreps
%\ba
%\chi_f (\bz) &=& \sum_{i=1}^N z_i \cr
%\chi_s (\bz) &=& \sum_{i=i}^N z_i^2 + \sum_{1=i<j}^N z_i z_j 
%= {1\over 2} \left(\sum_{i=1}^N z_i \right)^2
%+ {1\over 2} \sum_{i=1}^N z_i^2 \cr
%\chi_a (\bz) &=& \sum_{1=i<j}^N z_i z_j = {1\over 2} \left(\sum_{i=1}^N z_i \right)^2
%- {1\over 2} \sum_{i=1}^N z_i^2\\
%\chi_{ad} (\bz) &=& \sum_{i,j=1}^N z_i^{-1} z_j - 1 =  \left(\sum_{i=1}^N z_i^{-1} \right) 
%\left(\sum_{i=1}^N z_i \right) -1
%\nonumber\ea
\be
\label{lowreps}
\begin{split}
\chi_f (\bz) &= \sum_{i=1}^N z_i \ ,
\\
\chi_s (\bz) &= \sum_{i=1}^N z_i^2 + \sum_{j>i=1}^N z_i z_j 
= {1\over 2} \bigg(\sum_{i=1}^N z_i \bigg)^2
+ {1\over 2} \sum_{i=1}^N z_i^2 \ ,
\\
\chi_a (\bz) &= \sum_{j>i=1}^N z_i z_j = {1\over 2} \bigg(\sum_{i=1}^N z_i \bigg)^2
- {1\over 2} \sum_{i=1}^N z_i^2\ ,
\\
\chi_{ad} (\bz) &= \sum_{i,j=1}^N z_i^{-1} z_j - 1 =  \bigg(\sum_{i=1}^N z_i^{-1} \bigg) 
\bigg(\sum_{i=1}^N z_i \bigg) -1\ .
\end{split}
\ee
To obtain the above note that the singlet corresponds to $k_i=N-i$,  $i=1,\dots ,N$.  For the 
fundamental representation we  increase $k_1$ by one, leaving the other $k_i$'s intact.
For the symmetric representation we change $k_1$ by two and
for the antisymmetric representation we change $k_1$ as well as $k_2$ by unity.
Then, from \eqn{psar}, \eqn{slater} and \eqn{vanderm} we obtain the first three lines above.
The conjugate to the fundamental representation $\bar \chi_f$ is just $\chi_f$ with the $z_i$ replaced by their inverses. By decomposing 
$\chi_f \bar \chi_f = \chi_{ad} +1 $ we obtain $\chi_{ad} $ as above.
These will constitute basic building blocks when we examine the product of a large number of such irreps.

\subsection{Composition of fundamentals}

The composition of two irreps $r \otimes f$, one of which, specifically $f$, is the fundamental, is
particularly intuitive in the fermionic picture.
It corresponds to exciting one of the fermions in the state representing the original irrep by one unit of momentum.
This gives, in principle, $N$ resulting states, according to which of the $N$ fermions is excited. However,
excitations that would make two fermions occupy the same momentum level are impossible.

\no
It is easy to see that this reproduces the usual rules for Young tableau composition, adding a single box.
Adding one unit to $k_i$ amounts to increasing $\ell_i = k_i - k_N+i-N$ by one, except when $k_i = k_{i-1} -1$,
that is, $\ell_i = \ell_{i-1}$, in which case this is not possible.
Further, exciting the last fermion $k_N$ by 1 (when this is possible) amounts to subtracting 1 from all $\ell_j$,
that is, removing a column of length $N$ from the Young tableau.

\no
The fermion wavefunction of the state corresponding to $r \otimes f$ is, using \eqn{psis} and \eqn{lowreps}, simply
\be
\psi_{r \otimes f} (\bz) = \psi_{r} (\bz)\chi_f (\bz)= \psi_{r} (\bz) \sum_{i=1}^N z_i\ .
\ee
This can be used to obtain the fermionic state corresponding to the composition of several
($n$ in number) fundamental irreps $f$. The original, singlet state is simply $\Delta (\bz)$ and
an iteration of the above formula yields
\be
{\psi_{N,n}} (\bz) = \Delta (\bz) \Big(\sum_{i=1}^N z_i \Big)^n\ .
\label{Nspinia}
\ee
%where $ \Delta (\{ z_j \})$ is the Vandermonde determinant \eqn{vanderm}.
The irreducible components of the resulting representation can be "read off" from the terms of
$\psi_{N,n} (\bz)$ expanded as a polnomial in $z_j$: the coefficient of the terms
$z_1^{k_1} \cdots z_N^{k_N}$ such that $k_1 > k_2 > \dots > k_N$, represents the multiplicity
of the irrep corresponding to Young tableau lengths $\ell_j = k_j - k_N +j -N$, according to \eqn{ellk}.
The remaining terms, making $\psi_{N,n}$ properly fermionic, can be ignored. 

\subsection{Elementary example: Fundamental $SU(2)$ irreps}

As a simple example, we can apply the above method to the case of $SU(2)$ and make contact with previous
results. In this case \eqn{Nspinia} becomes simply
\be
{\psi_{2,n}} (z_1 , z_2 ) = (z_1 - z_2 ) (z_1 + z_2 )^n\ .
\ee
(In the above, and in similar formulas for $SU(N)$, we omit the phase prefactor of the Vandermonde
determinant  in \eqn{vanderm} as it only affects the $U(1)$ charge.)
Terms $z_1^{k_1} z_2^{k_2}$ with $k_1 > k_2$ can be isolated by writing $z_1 = z x$, $z_2 = z x^{-1}$,
which gives $z_1^{k_1} z_2^{k_2} = z^{k_1 + k_2} x^{k_1 - k_2}$. So powers of $z$ count total $U(1)$
charge, and focusing on terms $x^k$ for $k>0$ gives $k_1 - k_2 = k$. Clearly the single Young tableau
row length is $\ell = k_1 - k_2 -1= k-1 = 2j$, with $j$ being the spin of the irrep. We obtain
\be
\label{psiN2}
\begin{split}
{ \psi_{2,n}} & = z^{n+1} \left(x- x^{-1}\right) \left(x+ x^{-1}\right)^n 
\\
& = z^{n+1}\sum_{k=0}^n \binom{n}{k}  \left( x^{2k-n+1} - x^{2k-n-1} \right) 
\\
&= z^{n+1} \sum_{j=-n/2-1}^{n/2} {n!\, (2j+1) \over \left({n\over 2} -j\right)! 
\left({n\over 2} +j+1\right)!}\, x^{2j+1}   \ ,
\end{split}
\ee
with the sum running in integer decrements from $n/2$. The coefficient of the term 
$x^k = x^{\ell+1} = x^{2j+1}$ for $j \geqslant 0$
\be
\label{dkj} 
d_{n,j} = {n!\, (2j+1) \over \left({n\over 2} -j\right)! 
\left({n\over 2} +j+1\right)!}\ ,
\ee
is the multiplicity of spin $j$ in the decomposition, thus reproducing the result 
 of  \cite{Polychronakos:2016vpb}.\footnote{One may readily check that indeed
 $\displaystyle \sum_{j=j_{\rm min}}^{n/2} (2 j+1) d_{n,j}= 2^n$,
where $j_{\rm min}$ equals $0$ or $1/2$ depending on whether or not $n$ is even or odd.}

\no
A direct interpretation of (\ref{psiN2}) is obtained by focusing on the factor $(x+x^{-1})^n$.
Its expansion in $x^k$ produces the number of states $D_{n,k}$ at spin $S_z = k/2$.
Irreps with total spin $j\geqslant k/2$ contribute one state at $S_z = k/2$, so we can
isolate the number of irreps at $j=k/2$ by subtracting the contribution from $j\geqslant k/2 +1$, that is,
$D_{n,k+2}$. Given that $\ell = 2j$, we have
\be
\label{dlD}
d_{n,\ell} = D_{n,\ell}- D_{n,\ell+2} \ .%  D_{n,k-1}- D_{n,k+1}
\ee
which is precisely what the prefactor $x-x^{-1}$ is achieving, upon identifying $k=\ell+1$.

\no
The large-$n$, large-$j$ limit is interesting, and it is instructive to compute it in two alternative ways.  
For $n\gg 1$ and $j$ up to ${\cal O}(\sqrt{n})$, Stirling's formula
\be
\label{stilring}
n! \simeq \sqrt{2\pi}\, e^{-n} n^{n+\ha}\  ,\qquad n\gg 1\ .
\ee
applied to \eqn{dkj} yields 
\be
\label{fhj}
d_{n,j}\simeq {2^n \ov \sqrt{2\pi}} \Big({4\ov n}\Big)^{3/2} j \, e^{-2 j^2/n}\ ,
\ee
which coincides, for $j\gg 1$, with the result of  \cite{Curtright:2016eni,Polychronakos:2016vpb}. Note that the distribution \eqn{fhj}, for fixed $n$, has a maximum at $j=\sqrt{n}/2$ with  
a value of order ${\cal O}\big({2^n/n}\big)$.

\no
An alternative method is based on random walks, leading to a "diffusion process,"
which does  not rely on the exact formula for $d_{n,j}$ and is generalizable to higher groups. We focus on 
the factor $(x+x^{-1})^n$ in \eqn{psiN2} leading to $D_{n,k}$.
Each additional composition with a spin $\ha$ shifts the value of $k$ by either $+1$ or $-1$ and similarly the 
corresponding values for $\ell$. Explicitly,
\be
\label{dnl}
D_{n+1,k} =  D_{n,k+1} + D_{n,k-1} \ .
\ee
In our case $\displaystyle D_{n,k} = \binom{n}{n/2+k/2}$ which indeed satisfies \eqn{dnl}.
%and for $j=\ell/2$ yields \eqn{dkj}). 

\no
We will view \eqn{dnl} as a discrete evolution process and, for
large $n$ and $k$, we will take the continuum limit. To do this, we set $D_{n,k}$ to a
continuous function $D(n,k)$ end extract an explicit scaling factor $2^n$ to account for the doubling
of states per extra added spin
\be
\label{ton}
D(n,k) = 2^n F(n,k) \ ,\;~ \sum_k F(n,k) \simeq \int_{-\infty}^\infty dk F(n,k) = 1\ .
\ee
Then, a Taylor expansion of $F(n,k)$ in (\ref{dnl}) to leading orders in $n,k$ implies that this
obeys the heat equation
\be
2\, \del_n F = \del_k^2 F\ . 
\ee
Note that the extra factor $2^n$ in \eqn{ton} ensures that there is a smooth continuum limit of 
\eqn{dnl}, and $F(n,k)$ becomes a continuous distribution.
The solution of the above equation satisfying the initial condition $D_{0,k} = \delta_{0,k}$,
that is, $F(0,k) =\d(k)$, is given by 
\be
F(n,k) = {1\ov \sqrt{2 \pi n}} e^{-k^2/(2 n)}\ ,
\ee
for $n\geqslant 0$ and $k \in (-\infty, \infty)$, and thus
\be
\label{jhkjh11}
D(n,k) = {2^{n}\ov \sqrt{2\pi n }}  e^{-k^2/(2n)}\ .
\ee
Then (\ref{dlD}) becomes in this limit
\be
d (n,\ell ) \simeq  - 2\del_\ell D(n,\ell) = {2^{n+1}\ov \sqrt{2\pi n }} {\ell \over n} e^{-\ell^2/(2n)}\ .
\ee
Finally, changing variable to $j = \ell/2$ and taking into account that in the continuum limit $d_{n,j}$
becomes
a distribution, so that $d(n,j) dj = d (n,\ell ) d\ell = 2 d (n,\ell=2j ) dj$, giving an extra factor of $2$,
we reproduce \eqn{fhj}. This implies the proper normalization
\be
\int_{0}^{\infty}dj\, (2j+1) \  d(n,j) \simeq \int_{0}^{\infty}dj\, 2j \  d(n,j) = 2^n
\ee
where now the integration is over physical (positive) values of $j$,
with the multiplicity $2 j+1 \simeq 2j$ (valid for large $j$) included.

\subsection{Fundamental $SU(3)$ irreps}
\label{su3ir}

A more nontrivial case is that of the composition of $n$ fundamental $SU(3)$ irreps. 
%It is also instructive, in the sense that our results will be immediately generalizable to the $SU(N)$ case 
%that will be considered later in section \ref{suNir}.
For $SU(3)$, the fermion state \eqn{Nspinia} becomes
\be
\label{psin}
{\psi_{3,n }}= (z_1 - z_2)(z_1 - z_3)(z_2 - z_3) (z_1 + z_2 + z_3 )^n\ ,
\ee
(as in the $SU(2)$ case, we omitted the prefactor in the Vandermonde determinant
\eqn{vanderm}.)
To isolate terms $z_1^{k_1}\, z_2^{k_2}\, z_3^{k_3}$ with $k_1 > k_2 > k_3$ we adopt the parametrization
\be
z_1 = z x\, ,\quad   z_2 = z y\, ,\quad  z_3 = {z \over xy}\ ,
\ee
for which 
\be
\label{z1z2z3}
z_1^{k_1}\, z_2^{k_2}\, z_3^{k_3} = x^{k_1 - k_3}\, y^{k_2 - k_3}\, z^{k_1 + k_2 +k_3}
=x^{\ell_1 +2}\, y^{\ell_2 +1}\, z^{k_1 + k_2 +k_3}% = x^{m_1 +1} y^{m_2 +1} z^{k_3}
\ee
and focus on terms in the wavefunction with positive, ordered powers of $x$ and $y$. Clearly $z$
counts the $U(1)$ charge, so we can put $z=1$.
We also divide the wavefunction by $x^2 y$ to make its terms proportional to $x^{\ell_1} \, y^{\ell_2}$.
The reduced form of the wavefunction becomes
\be
\Psi_n = x^{-2}y^{-1}\left(x - y\right)\Big(x - {1 \over xy}\Big)\Big(y - {1\over xy}\Big)
\Big(x + y + {1\over xy} \Big)^n\ .
\ee
%Denoting by $\Psi |_{_+}$ positive power in $x$ and $y$ of $\Psi$ 
The generating function for the irrep content of the product of $n$ fundamentals of $SU(3)$
can then be expressed as
\be
\zeta_n (x,y) = \sum_{\ell_1 \geqslant \ell_2}\, d_{n;\ell_1 ,\ell_2}\, x^{\ell_1} y^{\ell_2} =
\left. 
{(x -y)(x^2 y-1)(x y^2 -1)\over x^4 y^3} \left(x+y +{1\over xy} \right)^n \right|_{+}\ ,
\label{zeta}
\ee
with $d_{n;\ell_1,\ell_2}$ the multiplicity of the irrep $\{\ell_1 , \ell_2 \}$ in the product, and
$|_{+}$ denoting that only positive, ordered powers $x^{\ell_1} \, y^{\ell_2}$ are kept.
As a demonstration, we list the nonvanishing $d_{n;\ell_1,\ell_2}$ for the first few values of $n$
\ba
& \zeta_0 = 1 \quad & d_{0;00} = 1\cr
& \zeta_1 = x\quad & d_{1;10} =1 \cr
& \zeta_2 = x^2 + xy \quad   & d_{2;20} = d_{2;11} = 1 \cr
& \zeta_3 = 1 + x^3 + 2x^2 y \quad  & d_{3;00} = d_{3;30} =1\, ,\ d_{3;21} = 2 \cr
& \zeta_4 = 3x + x^4 + 3x^3 y + 2 x^2 y^2 \quad  & d_{4;10} = d_{4;31} = 3\, ,\
d_{4;22} = 2\, ,\ d_{4;40} = 1\, , 
\label{fjh}
\ea
which reproduce the correct decomposition of $(\otimes f )^n$.
The coefficients $d_{n;\ell_1,\ell_2}$ for general $n$ can be calculated explicitly, but we defer their
derivation to the general $SU(N)$ case, where we will present a complete treatment.

\section{Composition of many fundamental $SU(N)$ irreps}
\label{suNir}

The fermion method described in the previous sections can be used to derive explicit combinatorial
expressions for the components of the decomposition of an arbitrary number $n$ of fundamental
$U(N)$ irreps. 
We will work directly with the fermion momenta $k_i$, leaving the transition to Young tableau row lengths 
$\ell_i $ in \eqn{ellk} for the end.

\subsection{Combinatorial expressions}

The fermion state for the product of $n$ $SU(N)$ irreps is  given by \eqn{Nspinia} and gives the
generating function of irreps in the momentum parametrization ${\bf k}$ (recall that we use boldface
letters for sets of $N$ quantities such as $k_1,\dots,k_N$)
\be
\psi_{N,n}(\bz)= 
\Delta (\bz) \Big(\sum_{i=1}^N z_i \Big)^n :=
\sum_{\bk}\, d_{n;{\bf k}}\, \prod_{i=1}^N z_i^{k_i} \ .
\label{psiNn}
\ee
The $\psi_{N,n}$ is antisymmetric in the $z_i$, and as a consequence the $d_{n;{\bf k}}$ appearing
above are fully antisymmetric in the $k_i$. When the $k_i$ are in decreasing order,
$d_{n;{\bf k}}$ gives the multiplicity of the irrep labeled by $k_1 > \dots > k_N$.

\no
To derive an explicit combinatorial expression for the multiplicity we first focus on
the coefficients produced by the term $(z_1+\cdots+z_N)^n$, denoted by $D_{n;{\bf k}}$.
We have
\be
\left(\sum_{i=1}^N z_i \right)^n := \sum_{k_1,\dots,k_N} D_{n;{\bf k}} \prod_{i=1}^N z_i^{k_i} 
= \sum_{k_1,\dots,k_N} \delta_{k_1 + \cdots +k_N ,n} \
{n \choose k_1,k_2,\dots,k_N} \prod_{i=1}^N z_i^{k_i} \ ,
\label{rann}\ee
that is,
\be
D_{n;{\bf k}}=
 \delta_{k_1 + \cdots +k_N ,n} ~ {n! \over \prod_{i=1}^N k_i !} \ , \qq  k_i \geqslant 0\ ,
\label{DNn}
\ee
where we used the standard expression for the multinomial coefficient.
The Vandermonde factor in (\ref{psiNn}) (again, omitting its phase prefactor)
contains $N!$ terms, each augmenting the powers of $z_i$ in
each term in the expansion. This corresponds to shifting the indices $k_1,\dots,k_N$ in the
$D_{n;k_1,\dots,k_N}$ summation by the opposite of the corresponding powers; that is,
\be
d_{n;{\bf k}} = \prod_{j>i=1}^N (S_i - S_j ) D_{n;{\bf k}}\ ,
\label{ranb}
\ee
with $S_i$ being discrete shift operators acting as
\be
S_i F(k_1,\dots,k_i,\dots,k_N ) = F(k_1,\dots,k_i -1,\dots,k_N )\ .
\ee
Substituting (\ref{DNn}) yields
\be
d_{n;{\bf k}} = 
 \delta_{k_1 + \cdots +k_N ,n+N(N-1)/2} \, n! \prod_{j>i=1}^N (S_i - S_j ){1 \over \prod_{r=1}^N k_r !}\ .
\label{shid}\ee
Observing that
\be
S_i {P(k_1,\dots,k_N ) \over \prod_{r=1}^N k_r !} = {k_i \, P(k_1,\dots,k_i -1,\dots, k_N )
\over \prod_{r=1}^N k_r !}\ ,
\ee
we see that for any polynomial function $P(k_1,\dots,k_N )$ the Vandermonde of shift operators $\prod_{j>i} (S_i - S_j )$ acting on the last term in (\ref{shid})
will produce a polynomial of degree $N(N-1)/2$ in the numerator with leading coefficient $1$.
Since it must also be fully antisymmetric in the $k_i$, the only such polynomial is the Vandermonde
of $k_i$. We obtain
\be
\label{lhq1}
d_{n;{\bf k}}= 
n!\, {\displaystyle \prod_{j>i=1}^N (k_i - k_j) \over \displaystyle \prod_{i=1}^N k_i !}\ ,
\quad\text{with}\quad \sum_{i=1}^N k_i = n+{N(N-1)\over 2}\ .
\ee
The above is written in the parametrization with a fixed total momentum.
To restore invariance under total momentum shifts $k_i \to k_i +c$,
we re-express the $k_i$ as
\be
\label{ksum}
k_i\to \, k_i +{n-k \over N}+{N-1\over 2}\, ,\quad  k := \sum_{i=1}^N k_i \ ,
\ee
which reduce to $k_i$ when the constraint $k=n+N(N-1)/2$ is satisfied, but otherwise are invariant
under the common shift $k_i \to k_i +c$, and thus $k \to k+ cN$. We obtain
\be
\label{lhq2}
d_{n;{\bf k}} = {\displaystyle n! \,\prod_{j>i=1}^N (k_i - k_j ) \over \displaystyle
\prod_{i=1}^N \Bigl(k_i +{n-k\over N}+{N-1\over 2}\Bigr)!}%\ ,\qq k = \sum_{i=1}^N k_i
\ .
\ee
The transcription in terms of Young tableau row lengths $\ell_i = k_i - k_N +i-N$ is immediate. Using
$\displaystyle \sum_{i=1}^N \ell_i = k - Nk_N -N(N-1)/2$ we obtain
\be
\label{dhjh}
d_{n; {\boldsymbol \ell}   }  = n! \, {\displaystyle \prod_{j>i=1}^{N-1} (\ell_i - \ell_j -i +j ) \,\prod_{i=1}^{N-1} (\ell_i +N-i)\over 
\displaystyle 
\Big({n-\ell \over N} \Big)! \, 
\prod_{i=1}^{N-1} \Bigl(\ell_i +{n-\ell \over N}+N-i\Bigr)!} \ ,\qq \ell := \sum_{i=1}^{N-1} \ell_i\ .
\ee
Formulae (\ref{lhq2}) and (\ref{dhjh}) hold whenever the $N$-ality condition 
$\ell \bb=\bb n\, (\text{mod}~N)$, 
or $k\bb=\bb n+ N(N\bb-\bb1)/2 \,(\text{mod}~N)$, is satisfied.

\no
The above multiplicities reproduce, as they should, the total number of states as
\be
N^n = \sum_{k_1>\cdots>k_N} \dim({\bf k} )\, d_{n;{\bf k}}\ ,
\label{dimsum}
\ee
where $\dim({\bf k} )$ is the dimension of irrep $\{k_1,\dots,k_n\}$, given by
\be
\dim({\bf k}) = \prod_{j>i=1}^N {k_i - k_j  \over j-i} 
= {\Delta({\bf k}) \over \prod_{s=1}^{N-1} s!}\ ,
 \label{dsk}
\ee
where $\D({\bf k})$ is the Vandermonde determinant  in  \eqn{vanderm} but without the prefactor.
To show this, we act on both sides of (\ref{psiNn}) with the Vandermonde of derivative operators
$\Delta({\bf \partial}) =\prod_{j>i}(\partial_i -\partial_j )$, where $\partial_i = \partial/\partial z_i$,
and then set all $z_i=1$.
On the left hand side, $\Delta({\bf \partial})$ acts only on $\Delta(\bz)$ (the action of any single
$\partial_i - \partial_j$ on the symmetric part gives zero).
%and, being a differential operator of order $N(N-1)/2$ acting on a polynomial of the same degree, it gives a constant.
The term $\partial_1^{n_1} \cdots \partial_N^{n_N}$ in 
$\Delta({\bf \partial})$, with $n_1,\dots,n_N$ a permutation of $0,1,\dots,N\bb-\bb1$, gives a nonzero result
only upon acting on the corresponding term $z_1^{n_1} \cdots z_N^{n_N}$ in $\Delta(\bz)$, equal to
$0!\, 1! \cdots (N-1)!$. Accounting for all $N!$ such terms we have
\be
\Delta({\bf \partial}) \Delta(\bz) = N! \prod_{s=1}^{N-1} s! = \prod_{s=1}^{N} s!
\ee
and thus
\be
\left. \Delta({\bf \partial}) \psi_{N,n}(\bz)\right|_{z_i =1}  = N^n \prod_{s=1}^{N} s!\ .
\label{DpDz}\ee
On the right hand side, we can antisymmetrize $\prod_i z_i^{k_i}$ into the full Slater determinant
(\ref{slater}), taking advantage of the antisymmetry of $d_{n;k_1,\dots,k_N}$, and restrict the summation
to $k_1 \bb>\bb \cdots \bb>\bb k_N$. Acting with $\Delta({\bf \partial})$ produces a symmetric polynomial
in $z_i$ with coefficients polynomials in $k_i$, and upon setting $z_i =1$
it gives an antisymmetric polynomial in $k_i$ of
degree $N(N\bb-\bb1)/2$, which is necessarily proportional to the Vandermonde $\Delta(\bk)$. 
By acting on the ground state $k_i = N-i$, in which case the Slater state is simply the Vandermonde
$\Delta(\bz)$, and using (\ref{DpDz}), the overall coefficient is found to be $N!$, for a result of
$N! \, \Delta(\bk)$. Equating left and right hand side results we obtain
\be
N^n \prod_{s=1}^{N} s! = \sum_{k_1>\cdots>k_N} d_{n;k_1,\dots,k_N} N! \,\Delta(\bk)\ ,
\ee
which, upon using \eqn{dsk}, reproduces (\ref{dimsum}).

\no
We note that the above procedure can be interpreted as a random walk on the
($N\bb-\bb1$)-dimensional lattice spanned by $x_i = k_i - k_{i+1}$, $i = 1,\dots, N-1$ with starting point
$x_1 \bb=\bb \cdots = x_{N-1} \bb=\bb 1$. Each
additional term in (\ref{rann}) increasing $n$ to $n+1$ can be interpreted as one of $N$ possible
steps on the lattice taking $x_{i-1} \to x_{i-1} -1 , x_i \to x_i +1$, $i=1,\dots,N$ ($x_0 = x_N :=0$),
and steps violating the boundary condition $x_i \geqslant 1$ forbidden. Imposing the boundary condition
can be achieved by leaving the walk unrestricted and summing over the $N!$ images of the starting
point obtained by permuting the initial $k_i$, each with a sign equal to the parity of the permutation.
Formula (\ref{ranb}) is implementing this image method: the term $D_{n;k_1,\dots,k_N}$ corresponds
to unrestricted walks starting from $x_1 = \cdots = x_N = 0$, and the $N!$ terms in the expansion
of the product of shift operators generates the images. This is a useful picture, especially in the
large-$n$ limit where the random walk essentially becomes a Brownian motion obeying a
diffusion equation, but we will not elaborate it further here.

\noindent
We conclude this section by pointing out that the number of $SU(N)$ irreps $d_{n;{\bf \ell}}$ can be interpreted
as an $N$-dimensional Catalan pyramid, just as the corresponding number of $SU(2)$ irreps maps to the
standard Catalan triangle numbers. 
%Specifically, for $N=2$, the Catalan triangle $C(\ell_1,\ell_2)$
%is defined as the number of words with $\ell_1$ letters $X$ and $\ell_2$ letters $Y$ such that no initial word
%has more letters $Y$ than letters $X$. This clearly reproduces the rules for composing 
%$n=\ell_1 \bb+\bb\ell_2$
%spin-$\half$ (fundamental, single-box) $SU(2)$ irreps resulting in a spin-$(\ell_1\bb-\bb \ell_2)/2$ irrep,
%$X$ representing boxes placed in the first row and $Y$ boxes placed in the second row.
We can define the $N$-dimensional Catalan pyramid $C(\ell_1 , \dots , \ell_N )$
as the number of words with $\ell_1$ letters $X_1$, $\ell_2$ letters $X_2$, \dots , $\ell_N$ letters $X_N$
such that no initial word has more $X_i$ than $X_{i+1}$ for all $i=1,2,\dots,N-1$ ($N=2$ gives the
usual Catalan triangle). This clearly
reproduces the rules of composing $n = \ell_1 \bb+\bb \ell_2 \bb+\bb \cdots \bb+\bb \ell_N$
$SU(N)$ fundamental irreps, $X_i$ representing boxes in row $i$, and maps to our $d_{n;{\bf \ell}}$.
We can also define truncated Catalan pyramids $C_{m_1,\dots,m_{N-1}} (\ell_1 , \dots , \ell_N )$
($0\leqslant m_{i+i} \leqslant m_i$),
generalizing Catalan trapezoids for $N=2$, as the number of words
such that in every initial segment of the word the number of $X_{i+i}$'s does not exceed the number of
$X_i$'s by more than $m_i$ ($m_i = 0$ reproduces the Catalan pyramid). These correspond to starting with an irrep of row lengths $m_i$ and adding
$n = \ell_1 \bb+\bb \cdots \bb+\bb \ell_N$ fundamental irreps for a final irrep of
row lengths $\ell_i + m_i$. The above Catalan structures also clearly admit an interpretation as the
number of random walks of the type described in the previous paragraph on the
$(N-1)$-dimensional integer lattice wedge $x_1 \geqslant x_2 \geqslant \cdots \geqslant x_{N-1} \geqslant 0$.

\subsection{Momentum density and a group duality}

We conclude by giving a "second quantized" expression for the $d_{n,k_1,\dots,k_N}$ that is useful
in the large-$N,n$ limit. Thinking of the $k_i$ as a distribution of fermions on the positive momentum
lattice $s=0,1,\dots$, we define the discrete momentum density of fermions $\rho_s$ equal to 1 on points
$s$ of the momentum lattice where there is a fermion and zero elsewhere; that is,
\be
\label{rs}
\rho_s = \sum_{i=1}^N \delta_{s,k_i}\ .
\ee
Clearly $\rho_s$ satisfies
\be
\sum_{s=0}^M \rho_s = N\ ,\qq \sum_{s=0}^M s\, \rho_s = k= n+{N(N-1) \over 2}\ .
\label{rNn}
\ee
In the above, $M$ is a cutoff momentum that can be chosen arbitrarily as long as it is
bigger than all the $k_i$.
Then (\ref{lhq1}) can be written as
\be
d_{n;k_1,\dots,k_N} = n!\, \prod_{t>s=0}^M (t-s)^{(\rho_s -1) \rho_t}\ ,
\label{discro}
\ee

\no
The integer $M$ could in principle be taken to infinity. However, keeping it finite serves to demonstrate an interesting
particle-hole duality of the formulae. Define
\be
{\tilde \rho}_s = 1-\rho_{M-s} ~, ~~~s=0,1,\dots,M\ .
\ee
Clearly ${\tilde \rho}_s$ is the density of holes on the lattice $[0,M]$ with the momentum reversed.
Moreover, ${\tilde \rho}_s$ satisfies
\be
\sum_{s=0}^M {\tilde \rho}_s = M-N+1\, ,\quad  \sum_{s=0}^M s\, {\tilde \rho}_s 
= n+{(M-N+1)(M-N) \over 2}\ .
\ee
Therefore, ${\tilde \rho}_s$ represents an irrep of $SU(M-N+1)$ with the same excitation $n$ (total
number of boxes) but with the
rows in the Young tableau of the $SU(N)$ irrep turned into columns for $SU(M-N+1)$, which defines the
dual irrep. One can check that
\be
d_{n;k_1,\dots,k_N} = n!\, \prod_{t>s=0}^M (t-s)^{({\tilde \rho}_s -1) {\tilde \rho}_t}\ .
\label{dualrho}
\ee
That is, in the decomposition of the tensor product of $n$ fundamentals of $SU(N)$, the multiplicity
of any given irrep is the same as the one for its dual irrep in the product of $n$ fundamentals of $SU(M-N+1)$.
Note that this relation holds for any $M$ such that $M \geqslant k_1 > k_2 > \cdots > k_N$.

\no
%Inspired by the physics literature we call this relation between groups and their representations a duality.
This duality between $SU(N)$ and $SU(M-N+1)$ can be turned into a self-duality if we choose $M=2N-1$,
which is possible if $k_1 \leqslant 2N-1$. This will be guaranteed, e.g., in the case $n <N$. Then (\ref{dualrho})
states that dual irreps in the decomposition come with equal multiplicities.

\no
The above group duality manifests as a particle-hole duality in the fermion density picture, but it can
also be understood in terms of Young tableau composition rules: when adding one additional box,
the rules are symmetric with respect to rows or columns, the only difference arising when a column
of length $N$ is formed, which is then eliminated. Therefore, starting with the singlet and adding
single boxes (fundamentals), the obtained set of irreps will be symmetric under row-column exchange
as long as $n<N$. This also shows that a similar duality will emerge in the composition of any
number of self-dual irreps, such as, e.g., $\ell_1 = 2, \ell_2 = 1$.

\subsection{Large-$n$ limit}

The limit $n \gg 1$ is similar to the large-$n$ limit in the $SU(2)$ case.
We can use either a Stirling approximation in the combinatorial formula for $d_{n;\ell_1 ,\dots, \ell_{N-1}}$
or $d_{n;k_1 ,\dots,k_N}$, or a random walk approach as in the $SU(2)$ case.
Either method leads to the desired result. Here we detail the derivation using the Stirling approximation 
\be
n! \simeq \sqrt{2\pi} n^{n+1/2} e^{-n} \ ,\qq n\gg 1\ ,
\ee
in the exact expression \eqn{lhq2} in which the total momentum is not fixed and shift invariance of the
$k_i$'s is manifest. Assuming $k_i \gg 1$, we obtain the intermediate expression towards a continuous
distribution $d_{n;k_1 ,\dots,k_N} \to d (n;{\bf k} )$
\be \label{fkh10}
d (n;{\bf k})\simeq {n^{n+1/2}\ov {\sqrt {2\pi}}^{N-1}}\,  e^{N(N-1)/2}  \prod_{j>i=1}^N (k_i - k_j ) 
\prod_{i=1}^N \bigg( k_i-{k\ov N} +{n\ov N} +{N-1\ov 2}\bigg)^{ -k_i+{k\ov N} -{n\ov N} -{N\ov 2}}\ .
\ee
In the $n\gg 1$ limit, a saddle point expansion in the product above reveals that $k_i - k/N$ scale as
$\sqrt n$. So, setting $\displaystyle x_i = {k_i\ov \sqrt{n}}$ and $\displaystyle x=\sum_{i=1}^N x_i$, we may write
%Performing this saddle point expansion and keeping the leading terms in $1/n$ gives
%\be
%\prod_{i=1}^N \bigg( k_i-{k\ov N} +{n\ov N} +{N-1\ov 2}\bigg)^{ -k_i+{k\ov N} -{n\ov N} -{N\ov 2}}
%\simeq  e^{-N(N-1)/2}\left({N\over n}\right)^{n+N^2/2} e^{-{N\ov 2n} \sum_{i=1}^N \big(k_i - {k\ov N}\big)^2}
%\ee
%Next, in the last product we pull out the factor $n/N$ resulting to the expression
\be
\label{fkh10}
d (n;{\bf k})\simeq {1\ov {\sqrt {2\pi}}^{N-1}}\,  {N^{n+N^2//2}\ov n^{(N^2-1)/2}}\, 
e^{N(N-1)/2}  \prod_{j>i=1}^N (k_i - k_j )\, J^{-1}_{{\bf k}}\ ,
\ee
where 
\be
J_{{\bf k}} = 
\prod_{i=1}^N \bigg(1 + {N\ov \sqrt{n}}\big( x_i-{x\ov N}\big) +{N(N-1)\ov 2 n}\bigg)^{ {n\ov N} + \sqrt{n} \big(x_i-{x\ov N}\big) +{N\ov 2}}\ ,
\ee
%where $\displaystyle x_i = {k_i\ov \sqrt{n}}$ is kept fixed in the large $n$ limit and $\displaystyle x=\sum_{i=1}^N x_i$.
Using the identity 
\be
\lim_{m\to \infty} \prod_{i=1}^N \Big(1+{\a_i\ov \sqrt{m}} + {\b_i \ov m} \Big)^{m+ \sqrt{m} \g_i} 
= \exp\Big(\sum_{i=1}^N\big( \a_i \g_i - \ha \a_i^2  + \b_i\big)\Big)\ ,
\ee
which holds provided that $\displaystyle \sum_{i=1}^N \a_i =0$, for the specific values
$\displaystyle m={n\ov N}$, $\displaystyle  \a_i=\g_i= \sqrt{N}\Big(x_i-{x\ov N}\Big)$ and
$\displaystyle \b_i= {N-1\ov 2}$, and taking the $n\gg 1$ limit  we obtain
\be
J_{{\bf k}} = e^{{N\ov 2} \sum_{i=1}^N \big(x_i - {x\ov N}\big)^2} e^{N(N-1)/2}
\ee
and \eqn{fkh10} becomes 
\be
d (n;{\bf k}) = {N^{n+N^2/2} \over \sqrt{2\pi}^{N-1} n^{(N^2 -1)/2}}  \prod_{j>i=1}^N (k_i - k_j ) \,
\small{\exp\Bigl[{-{N\over 2n}{\sum_{i=1}^N \left( k_i - k /N \right)^2}}\Bigr]}\ ,
\label{largenN}
\ee
an expression still invariant under uniform shifts of the $k_i$'s.
We see that in the large-$n$ limit the distribution of row lengths becomes Gaussian with a polynomial
prefactor, and typical row lengths of the obtained representations scale as $\sqrt n$.

\no
The above result can be expressed in terms of the dimension and quadratic Casimir of each irrep
$r = (k_1,\dots,k_N )$, given by
\be
\dim(r) = \prod_{j>i=1}^N {k_i - k_j  \over j-i}\ ,\qquad 
c_2 (r) = {1\over 2} \sum_{i=1}^N \left( k_i - {k\over N}\right)^{\bb 2} - {N(N^2 -1) \over 24}\ .
\ee
Using also
$\displaystyle  \prod_{j>1=1}^N(j-i)= \prod_{s=1}^{N-1} s!$, we obtain $d (n;{\bf k}):= d(N,n;r)$ as
\be
d(N,n;r) = \prod_{s=1}^{N-1} s! \,{N^{n+N^2/2}\, e^{N^2(N^2 -1) \over 24n} \over \sqrt{2\pi}^{N-1} n^{(N^2 -1)/2}}
\, \dim(r) \, e^{-{N\over n} c_2 (r) }\ ,
\label{dirr}
\ee
an expression reminiscent of the irrep-$r$ sector of the pure Yang-Mills partition function on the
Euclidean sphere.

\section{Coupling to magnetic fields}

As a physical application of the results of this paper, we examine the partition function of a system of
many fundamental $SU(N)$ "spins" (or better, flavors) coupled to an external non-abelian magnetic field.
This will require the evaluation of traces of exponentials of the total flavor operators in arbitrary irreps.

\no
Consider $n$ flavor multiplets, each transforming under the same irrep $R$ of
$SU(N)$, with generators $J_{s,a}$, $s=1,\dots, n$, $a=1,\dots, N^2-1$, in the presence of an external
nonabelian magnetic field $B_j$, $j=1,\dots N-1$ that couples to the Cartan generators of the flavors
$H_{s,j}$ according to the Hamiltonian
\be
H_{B,n}= -\sum_{s=1}^n \sum_{j=1}^{N-1} B_j H_{s,j} = -\sum_{j=1}^{N-1} B_j H_j\ ,
\label{HB}
\ee
where
\be
J_a = \sum_{s=1}^n J_{s,a} \ ,\quad H_j = \sum_{s=1}^{n} H_{s,j}\ ,
\ee
are the total $SU(N)$ operators and the corresponding Cartan components.
This is the $SU(N)$ analog of $SU(2)$ spins coupling to an external magnetic field through the unique
Cartan generator of the total spin $H = J_3$.
Note that this is the most general $SU(N)$ invariant coupling: any linear combination of the form 
$\sum_a B_a J_a$ can be brought to the form $\sum_h B'_j H_j$ by an appropriate $SU(N)$
conjugation (chance of basis in the flavor space), and if $B_a$ are common to all $n$ flavors this can
be achieved with a global $SU(N)$ rotation.

\no
If the full Hamiltonian is the one described by (\ref{HB}), the flavors
are not interacting and the full partition function is the $n^\text{th}$ power of the single-flavor
partition function
\be
{\cal Z}_n = {\cal Z}_1^n ~,~~~ {\cal Z}_1 = \tr_R\, e^{-\beta H_{B,1}} = \tr_R\, e^{\beta B_i J_i}\ .
\label{Zni}
\ee
So we would need to calculate the trace in (\ref{Zni}) only for the single irrep $R$. In general, however,
the flavors could couple. We will consider interactions that preserve the global $SU(N)$ symmetry and,
in fact, depend only on the total $SU(N)$ operators.
For instance, there could be an additional ferromagnetic-type interaction between the flavors,
similar to the ferromagnetic interaction of the $SU(2)$ spin case, for a full Hamiltonian of the form
\be
H = -C \sum_{a=1}^{N^2-1} J_a^2 - \sum_{j=1}^{N-1} B_j H_j\ ,
\ee
with $C$ a constant.
This extra ferromagnetic-type term is proportional to the quadratic Casimir $C_2$
of the total $SU(N)$. A more general interaction of this type would be any function $F(\{C_a\})$
of the full set of Casimirs $C_a$, $a=2,\dots,N$, of the total $SU(N)$
\be
H = F(\{C_a\}) -\sum_{j=1}^{N-1} B_j H_j\ .
\ee
In the presence of such an interaction the partition function does not factorize. Decomposing the full
flavor space into irreps $r$ of the total $SU(N)$, the partition function becomes
\be
{\cal Z} = \tr\, e^{-\beta H} = \sum_r d_{n;R,r}\, e^{-\beta F (r)} \,\tr_{r}
\exp \Bigl(\beta \sum_j B_j H_{j} \Bigr) 
\ee
with $\tr$ the trace in the full $n$-flavor Hilbert space, $F(r)$ the value of $F(\{C_a\})$ for the irrep $r$,
$\tr_r$ the trace in the irrep $r$, and $d_{n;R,r}$ the multiplicity of $r$ in the tensor product 
$(\otimes R)^n$. An evaluation of this partition function requires, apart from the multiplicity $d_{n;R,r}$,
the calculation of the trace $\tr_{r}\exp(\sum_j b_j H_j )$, where $b_j = \beta B_j$.

\no
The calculation of the trace $\tr_{r}\exp(\sum_j b_j H_j )$ for any
$SU(N)$ group and any irrep $r$ can be deduced from the matrix model analysis of section
\ref{generalities}. As stated there, energy eigenstates $\phi_{_E} (U)$  are labeled by irreps $r$
and have the form
\be
\chi_r (U) = \tr_r\, U\ .
\ee
The matrix $U$ can be written as
\be
U = \exp\bigl( i \sum_a \theta_a J_a \bigr) = V \exp\Bigl( i \sum_{j=1}^N x_j H_j \Bigr) V^{-1}\ ,
\ee
where $\theta_a$ are group parameters, $V$ is a diagonalizing matrix, and $H_j$ are the Cartan
generators, including the $U(1)$ part, in the diagonal basis with matrix elemens
\be
(H_j)_{ik} = \delta_{ij}\, \delta_{ik}\ .
\label{Hdiag}
\ee
Therefore, the energy
eigenstates provide the traces (characters)
\be
\chi_r (U) = \tr_r \exp\Bigl( i \sum_{j=1}^N x_j H_j  \Bigr)\ .
\ee
On the other hand, as explained in section \ref{generalities}, the energy eigenstates are given
as
\be
\chi_r (\bz) = {\psi_r (\bz ) \over \Delta (\bz)} \ ,\qq z_j = e^{i x_j}\ ,
\ee
with $\Delta (\bz)$ the Vandermonde determinant (\ref{vanderm}) and $\psi_r (\bz) $
a fermionic wavefunction given by the Slater determinant (\ref{slater}). Upon setting
$i x_j = b_j = \beta B_j$ we finally have
\be
\tr_r \exp\Bigl(\beta \sum_{j=1}^N B_j H_j  \Bigr) = {\psi_{\bk} (\bz) \over \Delta (\bz)} 
~,~~~z_j = e^{\beta B_j}\ .
\label{magtrace}
\ee
The irrep $r$ with Young tableau lengths $\ell_j$ maps to the fermionic momenta $k_j$ as
expressed in (\ref{ellk}).

\no
Note that,  (\ref{magtrace}) contains one extra magnetic variable, since there is also a term coupling to the
$U(1)$ charge of $U$ (the identity matrix). This can easily be eliminated by shifting all $k_j$
by a constant such that $k_1 + \cdots + k_N =0$ which neutralizes the $U(1)$ charge.
Alternatively, we can simply choose
magnetic parameters $B_j$ such that $B_1 + \cdots + B_j = 0$, ensuring that the coupling to
the $U(1)$ part vanishes. Further,
the magnetic variables $B_j$ appearing in the trace (\ref{magtrace}) couple to $H_j$ in the
specific basis (\ref{Hdiag}). Moving to any other basis, and in particular to one where the $H_j$ are 
traceless, would simply amount to replacing $B_j$ by appropriate linear combinations.

\no
The above traces constitute the basic components needed to determine the statistical mechanics of any
collection of flavors with Hamiltonian of the type considered in this section.

\section{Concluding remarks}

The $SU(N)$ decomposition problem is solvable in explicit form, at least for the direct
product of many "flavors" (fundamental irreps). The results in this paper open the road to various
applications and generalizations.

\no
Perhaps the most physically relevant question is the thermodynamics and phase transitions of a
large collection of $SU(N)$ flavors in the presence of an external nonabelian magnetic field. This is
in principle accessible from the results of the present work. For interactions of ferromagnetic type,
the system can be shown to exhibit ferromagnetic-like phase transition, but its phase structure and properties
turn out to be quite complicated and nontrivial, exhibiting qualitatively different behavior compared to the
standard $SU(2)$ ferromagnet. This system is treated in the follow-up paper \cite{Phases}.

\no
Another interesting issue, relevant to various theoretical physics contexts, is the scaling properties
of the system as the number of flavors $n$ and the size of the group $N$ both go to infinity.
The simplifications and special properties of the limit of large $N$ are known, and have been extensively
used in particle physics \cite{Hoo,BIPZ,GrMi}. In our case, the existence of the additional parameter
$n$ enriches the structure, and it
turns out that different limits can be reached according to the relation between $n$ and $N$ as they
both go to infinity. Again, in these limits phase transitions appear depending on the relation of
the two parameters. This  will also be investigated in a future publication.

\no
Finally, the decomposition of the direct product of irreps other than the fundamental can be considered. 
The relevant formulae quickly grow complicated and unwieldy (see \eqn{lowreps}), but the thermodynamic properties and
the various large-$n,N$ limits could still be analytically accessible, and are topics for future investigation.

%%%%%%%%%%
\subsection*{Acknowledgements}

We would like to thank Jules Lamers and Rafael Nepomechie for making us aware of related relevant
results in the literature, and the anonymous Reviewer for suggesting
the connection to Catalan numbers.\\
A.P. would like to thank the Physics Department of the U. of Athens for its hospitality
during the initial stages of this work. His visit was financially supported by a Greek Diaspora Fellowship
Program (GDFP) Fellowship.\\
The research of A.P. was supported in part by the National Science Foundation 
under grant NSF-PHY-2112729 and  by PSC-CUNY grants 65109-00 53 and 6D136-00 02.\\
The research of K.S. was supported by the Hellenic Foundation for
Research and Innovation (H.F.R.I.) under the ``First Call for H.F.R.I.
Research Projects to support Faculty members and Researchers and
the procurement of high-cost research equipment grant'' (MIS 1857, Project Number: 16519).

%\appendix

\end{document}